\documentstyle[epsfig]{mn} 
\newif\ifAMStwofonts

\def\eso36{\hbox{ESO~3.6-m}}                           
\begin{document} 
 
 \newif\ifAMStwofonts
 
\title[{Is the standard XRB picture in crisis?}] 
{ORIGIN OF THE X-RAY BACKGROUND AND AGN UNIFICATION: NEW PERSPECTIVES 
%
}  
 \author[A. Franceschini, et al.]  
{\parbox[]{6.5in}{A.\,Franceschini$^1$, 
V.\,Braito$^{1,2}$, D.\,Fadda$^3$ 
 }  
\\  
$^1$ Dipartimento di Astronomia, Universit\`a di  Padova, Vicolo Osservatorio 2,  
  I-35122, Italy; e-mail: franceschinini@pd.astro.it
\\  
$^2$ Osservatorio Astronomico di  Padova, Vicolo Osservatorio 5,  
  I-35122, Italy   
\\  
$^3$ Instituto de Astrofisica de Canarias, La Laguna, Tenerife, Spain\\ 
\\  
}

\maketitle  
  
  
\pagerange{\pageref{firstpage}--\pageref{lastpage}}   
  

\label{firstpage}  
  
\begin{abstract}  
We critically review the basic assumptions of the standard model for the 
synthesis of the XRB in the light of new data from ultradeep surveys by {\sl Chandra} 
and XMM, resolving major parts of it. 
Important constraints come in particular from the observed redshift distributions 
of faint hard X-ray sources -- showing large excesses at redshifts 
($z\sim 0.8$) much lower than expected by the {\sl synthesis models} --  
and from their X/optical/IR SEDs combined with the IR counts of type-II AGNs. 
We find that hard X-rays and the mid-IR appear to detect the same population of 
buried AGNs with peak emissivity around $z\sim 1$. 
This analysis, although supporting the general scheme which interprets the XRB  
as due to absorbed AGNs with broad N$_H$ distributions, requires major revision of 
the other postulate of the XRB {\sl synthesis models}:  the AGN unification. 
We argue that the unification scheme based on a simple orientation effect  
fails at high redshifts, where galaxy  
activity is induced by strong interactions and mergers among gas-rich systems. 
This helps explaining the observational evidence that type-I and II AGNs follow  
different evolutionary patterns, with type-I quasars providing a very biased trace of 
this activity.  
Combined deep X-ray and IR surveys consistently find that 
the universe has experienced a violent phase of galaxy activity around  
$z\simeq 1$, probably related with the assembly of massive galaxies. This has 
involved both star formation (primarily sampled in the IR) and  
obscured AGN fueling (as detected in hard X-rays and mostly responsible for
the XRB): our analysis implies that roughly 10 to 20\% of this activity 
has involved substantial AGN emission, this fraction likely reflecting the  
AGN/starburst duty cycle during the {\sl activation} phase. 

\end{abstract}  
  
\begin{keywords}  
Active galaxies:infrared, active galaxies:X-ray, AGN:surveys, galaxies:evolution,  
galaxies:active  
\end{keywords}

\section{INTRODUCTION}  
  
The X-ray Background (XRB) has been interpreted since long time as the integrated 
contribution of photo-electrically absorbed AGNs with a broad N$_H$ distribution and  
spread over a large redshift interval (Setti \& Woltjer 1989). A variety of 
observational tests, including X-ray spectral analyses and optical follow-up of very  
faint hard X-ray sources detected by {\sl Chandra} and XMM, have confirmed  
this assumption (Fiore et al. 2000; Hasinger et al. 2001;  
Barger et al. 2001; Alexander et al. 2001; Brandt et al. 2001).  
 
A very natural complement to this model was to postulate that the distributions of  
type I and II fractions and of the corresponding N$_H$ values
are ruled by the AGN unification scheme (Madau, Ghisellini \& Fabian 1994; 
Comastri et al. 1995). The ratio of obscured to unobscured objects was then taken 
to be $\sim 4:1$, as indicated by the statistics on  
local objects and consistent with the average covering factor of AGN tori (e.g.  
Maiolino \& Rieke 1995). 
Further assumption by these early XRB models is that these population distributions 
remain unchanged in the past, with the X-ray luminosity functions evolving back in 
time following the evolution laws observed for the type-I category (e.g. Miyaji 
et al. 2000): a strong increase of the X-ray emissivities from z=0 to z$\sim 2$ 
and a flattening thereof. 
 
With these assumptions, the predictive power of the {\sl synthesis models}  
was remarkable, as was the possibility to prove or disprove them. 
A first indication that some revisions of the above scheme should be considered has come 
from the very low fractions of luminous absorbed (type-II) quasars identified  
in faint X-ray samples (otherwise expected to be more numerous than the classical 
quasars). 
An attempt to adapt the XRB models to this new evidence was done by 
Gilli et al. (2001), by increasing the evolution rate of the type-II population 
above that of type-I. 
 
However, more radical revision of the {\sl synthesis models} seems now required by  
recent results of spectroscopic identifications of faint 
hard X-ray sources by {\sl Chandra} and XMM. 
At such faint limits substantial fractions of the XRB are resolved into sources. 
Rosati et al. (2002) and Hasinger (2002) show that the $z-$distribution for the absorbed
AGNs are very different from the model predictions. The statistics reported 
by Hasinger (2002) are rich ($\sim$300 redshifts) and complete (60\%) 
and come from the deepest hard X-ray surveys ever achieved: 
the majority of the sources are found at z$<1$, contrary to the expectations. 
 
\begin{figure}
\psfig{figure=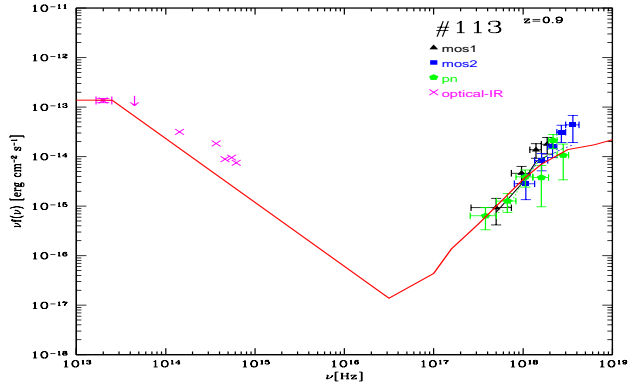,width=8.5cm,height=5.5cm}       
\caption{Broad-band spectrum of the source ISO J105306 +5728.2 
at z=0.9 detected by ISO and XMM in the Lockman Hole (source \#113 in Fadda et al. 2002). 
The XMM data are from Hasinger et al. (2001), analysed by Mucciarelli (2002). 
This source has the average $z$ and spectrum (N$_H\simeq 2.5\ 10^{23}\ cm^{-2}$)
of those in common ISO and XMM surveys, and is typical of type-II AGNs in general 
(Norman et al. 2002; Franceschini et al. 2000). 
The fitting curve corresponds to our adopted SED for type-II AGNs, redshifted 
to z=0.9 and normalized to the 15 $\mu$m flux.  
}  
\label{fig2}  
\end{figure}

We discuss in this Letter these unexpected results, also in the light of recent deep  
observations in the mid-IR revealing interesting properties of these faint X-ray 
sources found at moderate redshifts (Section 2). Section 3 compares the statistics 
of absorbed AGNs detected in hard X-rays and in the IR. Implications for  
the origin of the XRB and for the AGN unification are discussed in Section 4. 
 
The combination of very deep imaging in the IR and X-rays and the exploitation 
of the global constraints set by the X-ray and IR backgrounds offer for the first 
time the opportunity to strongly constrain the whole phenomenon of hidden  
gravitational accretion.
If the hard X-ray surveys sample the transmitted primary nuclear emissions,  
mid-IR surveys detect the re-radiated energy by the dusty circum-nuclear medium. 
A prediction of X-ray AGN counts based on the IR statistics is particularly useful 
if we consider that the population responsible for the XRB between 10 and 60 keV 
(where the bulk of the energy resides) cannot be presently observed in X-rays. 
We assume $H_0=50$ $Km~s^{-1}$ $Mpc^{-1}$, $\Omega_m=0.3$ and $\Omega_\Lambda=0.7$.

\section{RESULTS FROM RECENT COMBINED DEEP X-RAY AND IR SURVEYS} \label{sec:2}

Surveys in the mid-IR even at moderately deep flux limits have revealed remarkable  
capabilities to detect the IR counterparts of very faint hard X-ray sources. 
Hornschemeier et al. (2001) have correlated Chandra and ISO 15 $\mu$m images by 
Aussel et al. (1999) in the HDF-North and found that in the inner top-sensitivity region
6 out of the 8 {\sl Chandra} sources with $S_{2-10 keV}\geq 0.7\ 10^{-15}\ erg/s/cm^2$ 
are detected above $S_{15\mu}=100\ \mu Jy$. 
 
Combined XMM and ISO surveys in the Lockman Hole area have been 
published by Hasinger et al. (2001) and Fadda et al. (2002): here 63\% 
of the faint 5-10 keV sources are detected at 15 $\mu$m above 0.35 mJy. 
The large majority of these are clearly AGNs at 
0.5$<z<$1.5, as suggested by their high luminosities  
($L_{0.5-10 keV}\simeq 10^{42}-10^{45}$ erg/s).
Their optical colours and X-ray hardness ratios were used by Franceschini et al. (2002) 
to classify them as type-II or type-I AGNs: the observed relative fractions ($\sim$3:1) 
and bolometric luminosities were found 
consistent with the unified model, although this result may not be significant 
due to the poor statistics and the complex and uncertain selection biases. 
 
\begin{figure}
\psfig{figure=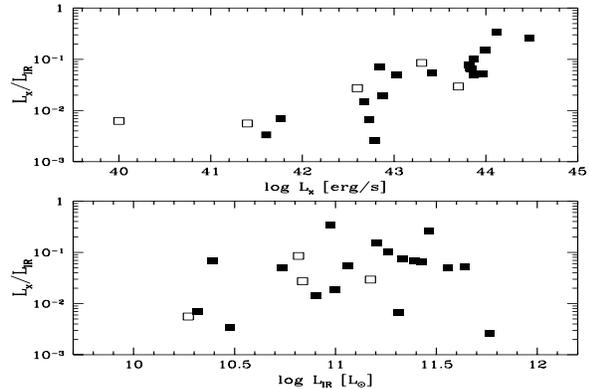,width=8cm,height=5.5cm}       
\caption{Top panel: ratios of the X-ray (2-10 keV) to 15 $\mu$  
luminosities as a function of the X-ray luminosity for ISO, XMM and {\sl Chandra} 
sources in the Lockman Hole (filled) and HDFN (open squares), where
$L_{x}=(5 keV)L_{5 keV}[erg/s/keV]$; $L_{IR}=(15\mu m)L_{15\mu}[erg/s/\mu m]$.
$L_{5 keV}$ is computed assuming a standard
power-law spectrum.       Bottom panel: same as top, vs. $L_{IR}$. 
}  
\label{fig1}  
\end{figure}

The optical-IR SEDs of type-II AGNs detected in the Lockman surveys and in the  
lensing cluster A2390 (Wilman et al. 2000; Franceschini et al. 2002) are consistent  
with these being completely buried AGNs (see one example in Fig. \ref{fig2}): 
the optical spectra reveal very red colours (often extremely red, R-K$>4$) which can  
be fit only as the redshifted emission by the host galaxy and {\sl 
show no scattered light from the AGN}. 
The unification model, which assumes that the AGN sequence is due to different 
line-of-sight inclinations of the torus axis in a randomly oriented sample, 
does not seem to apply to this population. 
 
Fadda et al. (2002) used these combined IR and X-ray survey data to quantify the 
relationship between the AGN emissions at the two wavelengths. The top panel in 
Fig. \ref{fig1} plots the ratios $L_x/L_{IR}$ of the 5 keV to 15 $\mu$ luminosities as 
a function of $L_x$ for sources in common in the HDFN and Lockman, and reveals a clear 
correlation: 
the less luminous X-ray sources have also very low flux ratios ($L_x/L_{IR}<0.01$). 
For these, both $L_x$ and the luminosity ratio are typical of  
starburst-dominated objects. 
On the contrary, the flux ratio appears not to depend on the IR luminosity  
(bottom panel).  
 
In summary, mid-IR observations appear to sample very efficiently AGNs with any 
levels of absorption and obscuration. 
 
\begin{figure}
\psfig{figure=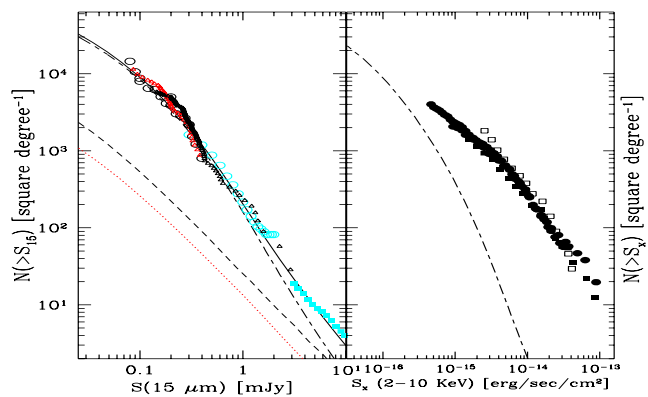,height=5.5cm,width=8.5cm}       
\caption{ 
Left panel: integral 15 $\mu$m counts from various ISO surveys (Elbaz et al. 1999),
with predictions of the evolution model by Franceschini et al. (2001). 
Dotted line: type-I AGNs; long-short dashes: 
evolving starbursts and type-II AGNs; dashes: non-evolving quiescent spirals;
continuous line: total. 
Right panel: integral 2-10 keV X-ray counts (Rosati et al.  
2002; Mushotzky et al. 2000; Hasinger et al. 2001), 
compared with our predicted contribution by type-II AGNs (long-short dashed line). 
} 
\label{cint1} 
\end{figure}

\section{AN ALTERNATIVE PICTURE FOR THE XRB} \label{sec:3}

Motivated by the fact that hard X-ray and IR samples provide the best  
complementary view of the AGN phenomenon (for both the transmitted primary emission  
and the dust-reradiated part), we exploit here our combined X/IR information 
to build an heuristic model for X-ray AGN evolution based on the
IR statistics. To this end, we assume that the types-I and types-II form two distinct
populations with independent evolutionary patterns, and concentrate here 
on modelling the type-II population which is of most interest.

A prediction for the X-ray luminosity function of the absorbed population
based on the IR data is made possible by the well-behaved relationship between 
15 $\mu$m and 2-10 keV fluxes (Fig. \ref{fig1}), where the $L_x/L_{IR}$ ratio, 
above the AGN threshold value of 0.01, shows moderate scatter and no dependence 
on $L_{IR}$. 
 
The Fadda's et al. (2002) analysis revealed that the AGN fraction (including type-I and 
II) in mid-IR selected samples is stable as a function of the IR flux around the value  
$f_{AGN}\simeq 20\%$ over a substantial flux interval between $S_{15\mu}\sim 3$  
mJy and $S_{15\mu}\sim 0.1$ mJy (while the bulk of the population are  
luminous starbust galaxies). 
This is a remarkable result, as this flux interval encompasses the bump in the 
counts (Elbaz et al. 1999) at $S_{15\mu}\simeq 0.3-0.5$ mJy which is dominated 
by the strongly evolving population of active IR galaxies (Franceschini et al.  
2001; Xu et al. 2001). 
A similar AGN fraction around 10\% was also found among high-z sub-mm SCUBA sources 
observed in X-rays (Fabian et al. 2000;  Hornschemeier et al. 2001; Barger et 
al. 2001).             {\bf  If the AGN fraction keeps so stable, this 
means that the AGN activity originating by gravitational accretion and the  
starburst activity of stellar origin should be tightly related with each other.} 
 
The modellistic analysis of the IR counts by Franceschini et al. (2001)  
identified three different evolutionary components:  {\sl (a)} non-evolving 
quiescent galaxies dominating the IR counts at the bright fluxes, {\sl (b)} 
evolving type-I AGNs as apparent in the UV-optical and soft X-ray quasar surveys, 
{\sl (c)} and a population of strongly evolving active galaxies including starbursts 
and type-II absorbed AGNs (these latter making up $\sim 15\%$ of the evolving  
population).  
The contributions of these various components to the integral 15 $\mu$m counts are 
reported in the left panel of Fig. \ref{cint1}.  
 
The model for the strongly evolving component, needed to reproduce the IR data
(multi-wavelength counts, the IR background, LLFs), was based on the IRAS LLF and 
assumed a strong increase with $z$ of the IR emissivity up to $z\simeq 0.8$ and a 
flattening thereof (we defer for all details on this model to Franceschini et al. 2001). 
 
\begin{figure}
\psfig{figure=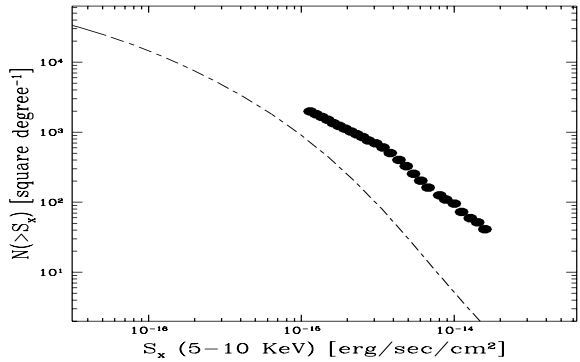,width=8cm,height=5cm}       
\caption{ 
Integral counts of X-ray sources selected in the 5-10 keV band (Rosati et al. 2002) 
compared with our predicted contribution by type-II AGNs. 
} %
\label{cint2} 
\end{figure}

Starting from this, assuming that a fraction $f_{AGN}=15\%$ of these active 
galaxies host optically buried but X-ray loud AGNs and adopting for these the  
broad-band SED as the fitting line in Fig. \ref{fig2}, it is straightforward to
estimate the contributions of these sources to the X-ray counts at various energies
(the X-ray K-corrections are also computed using this same hard X-ray spectrum in the
figure).
The 2-10 keV X-ray counts for type-II AGNs computed in such a way are reported in
the right panel of Fig. \ref{cint1}, while Fig. \ref{cint2} shows the 
predicted counts at harder energies (5-10 keV). 
These model counts are consistent with the (large) observed fractions  
($>50\%$) of X-ray sources with 15 $\mu$m counterparts reported by Hornschemeier 
et al. (2001) and Fadda et al. (2002). 
Note that our procedure does not account in detail for the observed spread in the 
X/IR flux ratio (Fig. \ref{fig1}), but we expect this would just moderately flatten 
the predicted counts.  

As apparent in Figs. \ref{cint1}b and \ref{cint2}, our expectation is that IR-selected 
type-II AGNs provide increasingly important contributions to the X-ray counts at 
increasing energies (also confirmed by the larger fraction at the harder energies
of identifications with faint IR sources, see Fadda et al. 2002).  
{\bf This implies that the counts are expected to become steeper at higher energies,  
in agreement with what observed by Rosati et al. (2002)}, while 
{\sl synthesis models} have difficulties in reproducing this observed trend.
 
\begin{figure}
\psfig{figure=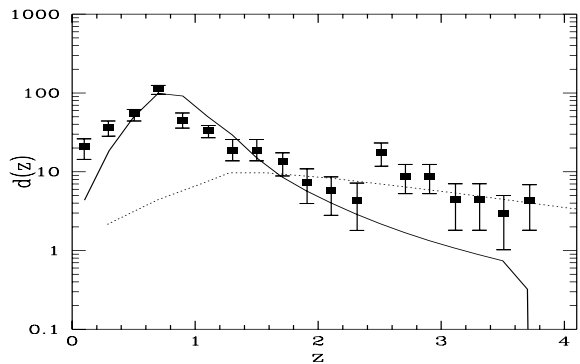,width=8cm, height=5.cm}       
\caption{ 
Redshift distributions from Hasinger (2002) of X-ray selected  
AGNs in deep $Chandra$ and XMM surveys with flux limits between  
$S_{2-10}\sim 5\ 10^{-16}$ and $\sim 1.5\ 10^{-15}$ erg/s/cm$^2$, respectively.  
This is a representative sample of the X-ray population at the faint flux limits. 
The dotted line is a prediction based on the XRB {\sl synthesis model} by Gilli et al. 
(2001) fitting the data only at $z>1.5$. 
The continuous curve is our predicted contribution of type-II AGNs assuming a flux 
limit of $S_{2-10}=8\ 10^{-16}$ erg/s/cm$^2$, and provides a remarkable description 
of the data at $z<2$, where the bulk of the XRB is produced. 
} 
\label{dz} 
\end{figure}

Our predicted contribution of type-II AGNs to the redshift distribution is  
reported in Fig. \ref{dz} (continuous line), providing a remarkably 
good fit to the data at $z<2$.   Type-I quasars, not included in our  
analysis, dominate the AGN population at $z>2$, as well as 
the counts at $S_{2-10}>$ few $10^{-15}$ erg/s/cm$^2$. 
The excess of observed sources at $z<0.4$ above the model prediction is due to the 
contribution of low-luminosity X-ray emissions by starburst galaxies, 
which are detectable at low redshifts. 
 
The contribution of type-II AGNs to the XRB spectrum is reported in Fig. \ref{xrb}.
This population provides $<50\%$ of the XRB below 5 keV, but dominates it above. 
The lower average redshift of these sources compared with those dominating the 
XRB in {\sl population synthesis} models allows in principle an easier fit to the 
XRB peak at 20 to 50 keV (for comparison, the synthetic 
spectrum e.g. by Gilli et al. [2001] appears shifted to slightly lower energies).

The time-dependent comoving volume emissivity of our model type-II AGN 
population is compared in Fig. \ref{cv} with that of type-I AGNs. The latter 
corresponds to the model adopted in Franceschini et al. (2001), and is consistent 
with results of quasar surveys in the optical and soft X-rays. 
The two clearly follow different evolutionary patterns.

\section{DISCUSSION AND CONCLUSIONS}\label{sec:discussion}

Not unexpectedly, the operations of a new generation of space observatories --
{\sl Chandra}, XMM-{\sl Newton} and ISO -- are dramatically improving 
our knowledge of the sources of the XRB. 
The current most popular view interprets the XRB and the X-ray source counts with 
evolutionary populations of absorbed and unabsorbed AGNs (the {\sl population 
synthesis models}), assuming that the relationship between the two is ruled 
by the AGN unification scheme and is constant with cosmic time. 
We have discussed in this Letter evidence that this improved knowledge of the XRB 
sources requires major revision of some of the {\sl population synthesis} assumptions.

\begin{figure}
\psfig{figure=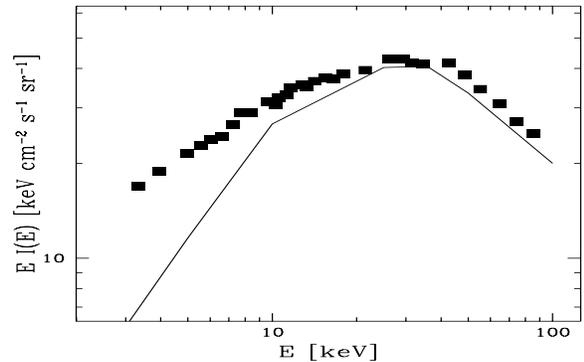,width=8cm, height=5.cm}       
\caption{ 
Contribution of the type-II AGN population to the XRB. Datapoints 
are from the A2 and A4 experiments on HEAO-1 (Gruber et al. 1999). 
} 
\label{xrb} 
\end{figure}

On one side, the new observations have confirmed the model's basic
postulate of the XRB at high energies as due to absorbed AGNs: these  
sources are indeed detected in hard X-rays and the hot dust emission 
associated with their absorbing medium in the mid-IR (Wilman  et al. 2000; 
Franceschini et al. 2002).  

\begin{figure}
\psfig{figure=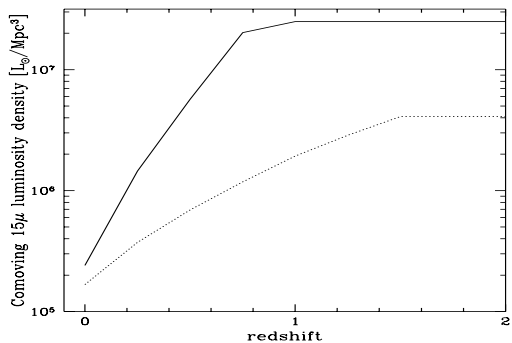,width=7cm, height=4.8cm}       
\caption{Comoving 15$\mu$m luminosity density as a function of $z$ for 
type-II (continuous line) and type-I (dotted line) AGN populations. 
} 
\label{cv} 
\end{figure}

On the other end, a revision of the model is required by the evidence that the 
evolutionary histories of type-I and type-II AGNs are very different. 
A critical observable in this sense are the redshift distributions  
of faint {\sl Chandra} and XMM X-ray sources, recently reported by Hasinger (2002). 
Although in these surveys the average identification fraction is $\sim$60\%, 
the observed $z$-distributions do not appear to vary appreciably with the X-ray flux 
or the optical magnitude limits. Hasinger's (2002) conclusion was that the 
distribution reported in Fig. \ref{dz} is 
representative of the whole population of faint X-ray sources responsible for 
the XRB. As illustrated in Fig. \ref{dz} and \ref{cv}, the outcome of these  
identifications is inconsistent with predictions of {\sl population synthesis} models.  
 
Other hints of a possible difficulty for the latter come from the shapes  
of the counts at various X-ray energies, showing evidence for steeper slopes 
at increasing energies, not completely reproduced by the {\sl synthesis models} 
(e.g. Rosati et al. 2002). 
Finally, the XRB spectral shape tends to appear harder than the 
{\sl population synthesis} predictions.

To explore new routes for the interpretation of the XRB and the X-ray AGN  
statistics, we have first relaxed the link between the type-I and type-II  
objects by assuming that the former evolve as found in optical and soft X-ray  
surveys, while the latter follow an independent evolutionary history. 
We have then modelled the type-II AGNs starting from the evolutionary model of  
active galaxies needed to reproduce multiwavelength IR counts (Franceschini et al. 
2001) and exploiting the evidence (Fadda et al. 2002; Fabian et al. 2000; Barger et  
al. 2001) that a constant fraction $f_{AGN}\sim 15\%$ of this active  
population contains buried AGNs. We have combined this information with the observed 
flux ratios between the 15 $\mu$m and X-ray fluxes, 
and with a typical hard X-ray spectrum as found in the deep  
XMM surveys (Fig. \ref{fig2}) for K-correction computations. 
 
With these assumptions, the expected integral counts of such hidden AGN population 
should continue steeply at faint X-ray fluxes, particularly at the high 
energies (Fig. \ref{cint2}). Correspondingly, these sources should dominate the 
XRB at $>10$ keV, while contributing a decreasing fraction at lower 
energies (Fig. \ref{xrb}).  
Due to the fast rise of the IR emissivity to $z\simeq 0.8$ and quick 
convergence thereof, the $z$-distributions of the faint hard X-ray sources 
are expected to have a maximum at this redshift of peak emissivity,  
hence explaining the outcomes of the spectroscopic surveys. 
 
We emphasize that these predictions rest on our only assumption that the type-II 
AGN fraction ($f_{AGN}\sim 15\%$), observed down to $S_{15\mu}=0.1$ mJy, 
does not drop immediately below. As a consequence, they should be quite robust. 
 
In conclusion, combined deep X-ray and IR observations consistently find that 
the universe has experienced a violent phase of galaxy activity around  
$z\simeq 1$, probably related with the assembly of massive galaxies. This
activity generates massive stars (whose flux is mostly re-radiated in the 
IR) and fuels nuclear BH's. 
Our analysis implies that roughly 10 to 20\% of this activity involves
substantial AGN emission mostly detectable as a faint hard X-ray flux. 
This fraction likely represents the duration of the AGN phase 
compared with that of the starburst during the {\sl activation} process. 
While their IR re-radiated part can only contribute a minor
fraction ($\sim 10\%$) of the cosmic IR background, their hard X-ray emission
is largely responsible for the XRB.
 
Type-I quasars turn out to be very biased tracers of this activity, their epoch
of maximal emissivity being quite displaced to $z\geq 2$. 
Absorbed and unabsorbed AGNs appear to form cosmologically distinct populations,
as anticipated in Franceschini et al. (1993). 
 
Consequently, the simple and attractive scheme trying to explain the XRB in the 
framework of the AGN unification scenario is ruled out by the present observations. 
The unification concept itself based on a simple 
orientation effect -- which works locally in the presence of modest amounts of  
ISM in regular rotationally-supported flows perturbed by weak  
galaxy interactions -- fails at higher redshifts. Galaxy activity during these past  
epochs is induced by strong interactions and mergers among gas-rich systems, which  
channel large amounts of absorbing gas around the nuclear source, completely  
covering it. 
 
 
The different evolutionary patterns observed for type-I and II AGNs require 
an explanation.
There may be an effect induced by luminosity, favouring the optically-thin 
quasar phase in relation with the highest-mass nuclear BH's, quickly expelling 
the surrounding gas by radiation pressure. 
Then the two AGN categories may trace environments with different cosmic densities:
type-I AGNs tracing the highest density peaks which evolve on faster cosmological 
timescales, and type-II AGNs, present in lower density concentrations, 
showing protracted activity down to much lower redshifts.  
Once again, significant complication of previous schemes and new imagination are 
required by new observational facts.

\section*{Acknowledgments}               
Work supported by the EC RTN Network "POE" (contract HPRN-CT-2000-00138) and the 
Italian Space Agency. 
We would like to thank X. Barcons, R. Gilli, G. Hasinger, G. Risaliti and P. Tozzi 
for useful comments and discussions.

\end{document}